\documentstyle[epsfig,prl,aps]{revtex}  
\begin{document}


 \twocolumn[\hsize\textwidth\columnwidth\hsize  
 \csname @twocolumnfalse\endcsname              

\title  {
Boson Realization of the SU(4) Model of  
High-Temperature Superconductivity
}

\author  {J. Dobe\v{s}
         }
\address  {
                     Nuclear Physics Institute,
                  The Academy of Sciences of the Czech Republic,
                     CS 250 68 \v{R}e\v{z}, Czech Republic
          }

\date{\today}

\maketitle


\begin{abstract}
The SU(4) algebraic model of high-temperature superconductivity is
studied employing the boson mapping techniques.  The bosonization of
the model enables us to get exact numerical solution of the model.  The
order parameters are discussed.  The situation close to the SO(5)
dynamical symmetry limit is interpreted as a modelling case for the
behaviour of $D$-wave superconducting and antiferromagnetic phases in
cuprates.

\end{abstract}


 ]  

\narrowtext


The high-temperature superconductivity represents a challenging task
for the theory of many-electron correlated systems. Data suggest that
it is related to the singlet D-wave pairing (dSC) phase. One of the
intriguing aspects is then a proximity of the dSC phase to the
antiferromagnetic (AF) phase. Indeed, in cuprates, the AF order is
observed at half filling whereas the dSC order develops as the material
is doped by holes.

The interplay between the dSC and AF phases is treated by employing the
symmetry principles in the SO(5) model proposed by S.-C.~Zhang
\cite{Zhang}. Two dSC order parameters and three AF order
parameters are unified in a five dimensional vector. The SO(5) rotations 
induce then transitions between two phases.  Microscopic models
exhibiting the SO(5) symmetry have also been discussed
\cite{Rabello,Henley}.

Recently, Guidry and collaborators have embedded the SO(5) algebra
into a larger SU(4) algebra \cite{Guidry,Wu}.  The microscopic
realization of the SU(4) generators has been given and the three
dynamical symmetry limits of the SU(4) model have been studied. The
SO(5) limit is interpreted as a critical dynamical symmetry
interpolating between the dSC and AF phases.

In the present paper, we discuss the boson realization of the SU(4)
algebra. The bosonization of the model enables us to get exact
numerical solution of the model. The closed forms for the dSC and AF
order parameters in the dynamical symmetry limits are obtained. The
interplay between two phases is studied also outside these limits.  In
agreement with Ref.\cite{Wu}, the situation close to the SO(5) limit is
suggested as a modelling case to the real behaviour in cuprates.

The 15 SU(4) generators are microscopically constructed on
a lattice with the degeneracy $\Omega$ as
\cite{Guidry}

\begin{eqnarray}   \label{SU4ope}
D^{\dagger} & = &\sum_{\bf k} g({\bf k}) c^{\dagger}_{{\bf k},\frac{1}{2}}
c^{\dagger}_{-{\bf k},-\frac{1}{2}}
\nonumber \\
D & = & (D^{\dagger})^{\dagger}  
\nonumber \\
\pi^{\dagger}_{\mu} & = & \frac{1}{\sqrt{3+(-)^{\mu}}}
\sum_{{\bf k},i+j=\mu}  g({\bf k}) c^{\dagger}_{{\bf k}+{\bf Q},i}
c^{\dagger}_{-{\bf k},j}, \;  \; \;
\mu=-1,0,1  
\nonumber   \\
\tilde{\pi}_{\mu} & = & (-)^{1-\mu} (\pi^{\dagger}_{-\mu})^{\dagger}      
\nonumber \\
S_{\mu} & = & \frac{1}{\sqrt{3+(-)^{\mu}}}
\sum_{{\bf k},i-j=\mu} {\rm sgn}(j) c^{\dagger}_{{\bf k},i} c_{{\bf k},j}, \;  \; \;
\mu=-1,0,1   
\nonumber  \\ 
Q_{\mu} & = & \frac{1}{\sqrt{3+(-)^{\mu}}}
\sum_{{\bf k},i-j=\mu} {\rm sgn}(j) c^{\dagger}_{{\bf k}+{\bf Q},i} 
c_{{\bf k},j}, \;  \; \;
\mu=-1,0,1   
\nonumber  \\ 
n & = & \sum_{{\bf k},i} c^{\dagger}_{{\bf k},i} c_{{\bf k},i} \; \; .
\end{eqnarray}

Here, $c^{\dagger}_{{\bf k},i}$ 
denotes the creation operator of a fermion with
momentum ${\bf k}$ and spin projection  $i=-\frac{1}{2}, \frac{1}{2}$,
$\bf Q$ is an AF ordering vector 
so that $\exp ({\rm i} {\bf Q} \cdot {\bf x})= \pm 1$ at every
lattice site, and $g({\bf k})={\rm sgn} (\cos \, k_x -  \cos \, k_y)$
with the properties $g({\bf k}+ {\bf Q}) = - g({\bf k}) $
and  $|g({\bf k})|=1$.\cite{foot1,foot2}

There are three dynamical symmetry chains 
conserving total spin $S$ and particle number  $n$
(or charge $M=\frac{1}{2} (n-\Omega)$)
discussed in \cite{Guidry,Wu}: \\
i) the SO(4) limit comprising a subset of operators of staggered magnetization  
$Q_{\mu}$ and spin $S_{\mu}$, \\
ii) the SO(5) limit with the triplet $\pi$-wave creation and
annihilation operators $\pi^{\dagger}_{\mu}$ and $\tilde{\pi}_{\mu}$,
and  $S_{\mu}$, \\
iii) the SU(2) limit with the singlet $D$- wave creation and
annihilation operators $D^{\dagger}$ and $D$ (and $S_{\mu}$).

One can move between the dynamical symmetry limits with a Hamiltonian
\begin{equation} \label{HI}
H_{{\rm I}} = -\frac{G_{{\rm I}}}{2} ((1+x) D^{\dagger} D 
-(1-x) \pi^{\dagger} \cdot \tilde{\pi})
\end {equation}
with $x=1$ for the SU(2) limit,
$x=-1$ for the SO(5) limit,
and $x=0$ for the SO(4) limit.
The SU(4) quadratic Casimir operator
$$
C_{{\rm SU(4)}} =
D^{\dagger} D - \pi^{\dagger} \cdot \tilde{\pi}
+ Q \cdot Q +  S \cdot S +M(M-4)
$$
gets for the most symmetric collective SU(4) subspace
(which is only discussed in the present paper)
the value $\frac{1}{4} \Omega (\Omega+8)$.
Then, an alternative parameterization of the SU(4) Hamiltonian
up to uninteresting for the present discussion constant, charge, 
and quadratic spin  terms
is written as \cite{Guidry,Wu}
\begin{equation}  \label{HII}
H_{{\rm II}} = -G_{{\rm II}} ((1-p) D^{\dagger} D 
+p Q \cdot Q )
\end {equation}
with
$p=0$ for the SU(2) limit,
$p=0.5$ for the SO(5) limit,
and $p=1$ for the SO(4) limit. 

Even though the Hamiltonians (\ref{HI}) and (\ref{HII})
must generally be equivalent, the assumed range of the
parameters $G_{{\rm I}}, G_{{\rm II}} \geq 0$, $1 \geq x \geq -1$,
and $1 \geq p \geq 0$   leads to  different
physical picture.  For the SU(2) limit, both parameterization 
agree. For the SO(4) and SO(5) limits and in-between, however,
the positive values of the interaction strengths 
$G_{{\rm I}}$ and $G_{{\rm II}}$ imply different signs of the
Hamiltonians (\ref{HI}) and (\ref{HII}). The spectra of two
parameterization are reversed with the ground state of 
(\ref{HI}) being the highest-lying state of (\ref{HII})
and {\em vice versa}.

For the dynamical symmetry limits, the SU(4) Hamiltonian is
analytically solvable. Outside the limits, the numerical
diagonalization must be performed. In both these tasks, 
an application of the 
boson mapping techniques proves to be useful.  The Dyson boson
realization of the SU(4) algebra is constructed by mapping the
bifermion operators (\ref{SU4ope}) onto bosonic operators
formed from the scalar boson operators $d$ and
vector boson operators $p$ \cite{foot3}
\begin{eqnarray} \label{bosmap}
D^{\dagger} & \rightarrow & (\Omega + 2 - n) d^{\dagger} + 
(d^{\dagger}d^{\dagger} - p^{\dagger} \cdot p^{\dagger} ) d  
\nonumber \\
D & \rightarrow & d 
\nonumber \\
\pi^{\dagger}_{\mu} & \rightarrow & (\Omega + 2 - n) p^{\dagger}_{\mu} + 
(d^{\dagger}d^{\dagger} - p^{\dagger} \cdot p^{\dagger} ) 
\tilde p_{\mu} 
\nonumber  \\
\tilde \pi_{\mu} & \rightarrow & \tilde p_{\mu} 
\nonumber \\
Q_{\mu} & \rightarrow & p^{\dagger}_{\mu} d - d^{\dagger} \tilde
p_{\mu}
\nonumber \\
S_{\mu} & \rightarrow & \sqrt{2} [p^{\dagger} \tilde p]^{1}_{\mu}  
\nonumber \\
n & \rightarrow & 2 (d^{\dagger} d - p^{\dagger} \cdot \tilde p)
\; \; .
\end{eqnarray}


By employing (\ref{bosmap}), one can easily construct the boson image
of the SU(4) Hamiltonian.  The bosonized task is relatively simple to
solve numerically in the boson space SU(1)$_d \otimes$SU(3)$_p$
\cite{foot4}.
To obtain the spectrum of $S=0$ states including the
ground state, it is sufficient to diagonalize a matrix of dimension $n/2$.
That represents an enormous truncation of the original fermion space.

In the boson treatment, we obtain the exact and full solution
of the fermion problem. The eigenstates states have 
good charge and spin quantum numbers which reflect the
symmetry of the original task.
For them, we relate the dSC order parameter to the
matrix element of the $D$-pair transfer between the  ground states
$$
\alpha_0 = \langle n+2 | D^{\dagger} | n \rangle 
\; ,
$$
whereas the AF order parameter is connected to the
reduced matrix element of the $Q$ operator  between the
$S$=0 ground state and the lowest lying $S$=1 state 
\cite{foot5}
$$
\beta_0 = \langle n  S=1 || Q || n  S=0 \rangle 
\; .
$$

Even if the $D^{\dagger}$
and $Q$ order operators do not
always belong to the
generators of the particular dynamical symmetry
limit chains, we have succeeded to obtain the analytical
formulas by inspecting the numerical results.
We have got
\begin{eqnarray}   \label{orderSU2}
{\rm  SU(2)} & &  {\rm parameterizations \; (\protect\ref{HI}) \; and 
\; (\protect\ref{HII})} 
\nonumber \\
\alpha_0^2 & = & \frac{1}{4}(n+2)(2\Omega-n)                  
\nonumber \\
\beta_0^2 & =& \frac{3}{4} \frac{2\Omega-n}{\Omega-1}n
\; ,
\end{eqnarray}
\begin{eqnarray*}
& & {\rm SO(5)} \; \;  {\rm parameterization \;(\ref{HI}) }  
\nonumber  \\
\alpha_0^2& =&\frac{(2\Omega-n)(2\Omega-n+2)}{4(\Omega+1)}   
\; \; \; {\rm for \;} n=4k
\nonumber  \\
\beta_0^2& =&\frac{2\Omega-n}{4(\Omega+1)}n
\; \; \; {\rm for \;} n=4k
\; ,
\end{eqnarray*}
\begin{eqnarray}  \label{orderSO5}
& & {\rm SO(5)} \; \;{\rm parameterization \;(\ref{HII})} 
\nonumber  \\
\alpha_0^2& = &\frac{1}{4} \frac{\Omega-n}{\Omega-n+3}
(n+2)(2\Omega-n+6) 
\nonumber    \\
\beta_0^2& =&\frac{3}{4} \frac{2\Omega-n+8}{\Omega-n+5} n
\; ,
\end{eqnarray}
\begin{eqnarray*}
& & \; \; \; \; 
{\rm SO(4)} \; \; {\rm parameterization \;(\ref{HI})}
\nonumber  \\
\alpha_0^2& = &\frac{1}{16} (2\Omega-n+6)(n+2) 
\; \; \; {\rm for \;} n=4k+2     
\nonumber   \\
\beta_0^2& =& 0 \; \; \; {\rm for \;} n=4k
\; ,
\end{eqnarray*}
\begin{eqnarray} \label{orderSO4}
{\rm SO(4)} &\; \;& {\rm parameterization \;(\ref{HII})}
\nonumber \\
\alpha_0^2 &= &\frac{1}{4} (\Omega-n)(n+4) 
\nonumber \\
\beta_0^2& =& \frac{1}{4} n (n+4)
\; .
\end{eqnarray}

The results (\ref{orderSO5}) and (\ref{orderSO4}) for the SO(5) and SO(4)
limits of the parameterization (\ref{HII}) are remarkable.
In the SO(4) limit, the AF order parameter $\beta_0$ increases linearly with
$n$ up to the value $\Omega/2$ for $n=\Omega$.
The dSC order parameter $\alpha_0 \rightarrow 0$ when
$n \rightarrow 0$ and $n \rightarrow \Omega$, and reaches the maximum
value $\frac{1}{4} \Omega$ at the quarter filling $n=\Omega /2$.

In the SO(5) limit, the order parameters for $n \ll  \Omega$ agree
with the values of the SC case (\ref{orderSU2}) since then the SO(5)
ground state of the parameterization (\ref{HII}) has the form of the
D-pair condensate $ | {\rm g.s. } \rangle \propto D^{\dagger \;
\frac{1}{2} n}  | 0 \rangle $. The situation changes
for $n \rightarrow \Omega$ when a steep increase appears in the AF order
parameter whereas the SC order parameter goes to zero. One may get more
insight  into this phase transition by introducing
the fractional doping of holes $\delta = 1 - n /\Omega$. Then
for $\Omega \rightarrow \infty$, Eqs.(\ref{orderSO5}) are rewritten as
\begin{eqnarray*}
\alpha_0 / \Omega & = & \frac{1}{2}
\left[ \frac{\Omega \delta}{\Omega \delta +3} (1-\delta ^2)
\right]^{\frac{1}{2}}
\nonumber
\\
\beta_0 / \Omega & = &\frac{1}{2}
\left[ \frac{3}{\Omega \delta +5} (1-\delta ^2)
\right]^{\frac{1}{2}} 
\; ,
\end{eqnarray*}
from which form the difference between the half filled case  ($\delta=0$)
and the hole doped situation ($\delta>0$) is easily seen.

\begin{figure}[thb]
\epsfxsize 8.5cm
\centerline{\epsfbox{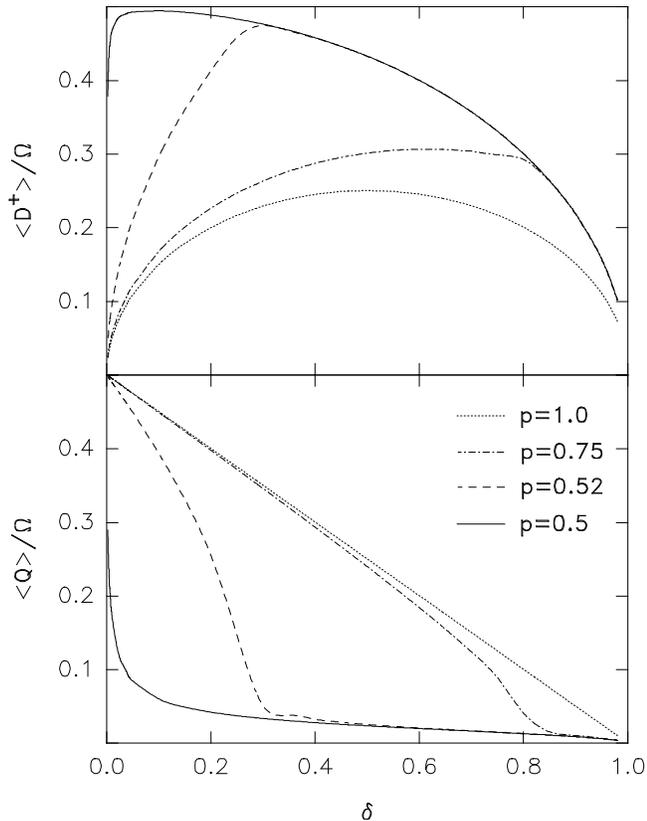}}
\caption{AF order parameter (bottom panel) and
dSC order parameter (top panel) as a function of the
fractional doping of holes $\delta$. Degeneracy of the laticce
$\Omega$=2000. 
Curves are shown for the values of the parameter $p$ in
(\protect\ref{HII})   $p$=1 (SO(4) limit), 
$p$=0.5 (SO(5) limit) and two intermediate cases
$p$=0.52 and $p$=0.75.}
\label{fig1}
\end{figure}

In Fig.\ref{fig1}, the order parameters are shown as obtained 
with Hamiltonian (\ref{HII}) for
$\Omega=2000$. Besides the dynamical symmetry limits SO(5) ($p$=0.5)
and SO(4) ($p$=1), two calculations are given for the
values of the parameter $p$=0.52 and 0.75. 
Outside the dynamical symmetry limits, the order parameters are close
to the SO(5) values for $\delta \rightarrow 1$
and to the SU(4) values for $\delta \rightarrow 0$. The results for the
AF order parameter in the case $p$=0.52 are very much similar to those
obtained in Ref.\cite{Wu} with the coherent state approach.

We thus confirm the findings of Refs.\cite{Guidry} and \cite{Wu}
that the SU(4) algebraic model in a situation close to its
SO(5) dynamical symmetry limit is a convenient tool to model
behaviour of high-temperature-superconducting cuprates. The transition
from the AF phase to the dSC phase naturally appears as the 
system is doped by holes starting from the half-filled case.

Attractive feature of the SU(4) algebraic model is its exact
solvability. Then despite the fact that some portion of the real
physics might not be present in the model (there is no kinetic-energy
term in the SU(4) Hamiltonian), it can still be useful in testing
approximate many-body procedures. For the dynamical symmetry limits we
have obtained closed expressions from which behaviour of the order
parameters in the phase transition region is seen explicitely.


This work has been supported by the Grant Agency of the
Czech Republic under grant 202/99/0149. 

\baselineskip = 14pt
\bibliographystyle{unsrt}

\begin{thebibliography}{99}
\bibitem{foot1} Our $\pi^{\dagger}$ operator differs by a
multiplicative factor i from $\pi^{\dagger}$ of Ref. \cite{Guidry}.
\bibitem{foot4}
Of course, an alternative methods for the numerical solution may 
employ the algebraic properties of the
fermionic SU(4) algebra and any of its dynamical symmetry chains.
\bibitem{foot5} In the dynamical symmetry limits both in the
parameterizations (\ref{HI}) and (\ref{HII}),
the first excited $S$=1 state exhausts fully the $Q$-strength from the
ground state. Similarly, the $D$-pair transfer from the ground state
goes fully or almost fully to the ground state of the adjacent
system. An exception is the SO(5) limit in the parameterization
(\ref{HI}) for $n=4k+2$ case.
\end{thebibliography}

\end{document}